\documentclass[aps,twocolumn,showpacs]{revtex4}
\usepackage{psfig}

\def\kbar{\protect\@kbar}
\def\@kbar{%
\relax \bgroup
\def\@tempa{\hbox{\raise.73\ht0
\hbox to0pt{\kern.25\wd0\vrule width.5\wd0
height.1pt depth.1pt\hss}\box0}}%
\mathchoice{\setbox0\hbox{$\displaystyle k$}\@tempa}%
{\setbox0\hbox{$\textstyle k$}\@tempa}%
{\setbox0\hbox{$\scriptstyle k$}\@tempa}%
{\setbox0\hbox{$\scriptscriptstyle k$}\@tempa}%
\egroup}

\input epsf

\begin{document}

\title{A universal ionization threshold for strongly driven Rydberg states}
\author{Andreas Krug\footnote{present address: inuTech GmbH, F\"urther Str. 212, D-90429
N\"urnberg} and Andreas Buchleitner}
\affiliation{Max-Planck-Institut
f\"{u}r Physik komplexer Systeme, N\"{o}thnitzer Str.\ 38, D-01187 Dresden} 

\date{\today}
\begin{abstract}
We observe a universal ionization threshold for microwave driven one-electron
Rydberg states of H, Li, Na, and Rb, in an {\em ab initio} numerical treatment
without adjustable parameters. This sheds new light on old experimental data,
and widens the scene for Anderson localization in light matter interaction.
\end{abstract}

\maketitle

The suppression of quantum transport across disordered media 
is one of the most spectacular consequences of destructive 
quantum interference. Originally predicted by Anderson
\cite{anderson58} 
in his
treatment of electrons propagating in disordered one dimensional
lattices, {\em Anderson localization} has now become a general concept which 
prevails in abundant scenarios of coherent quantum transport in the
presence of disorder \cite{wiersma97,kramer93}. Once it was realized that
dynamical chaos has 
the potential to substitute disorder in the long time evolution of 
low dimensional systems on the classical as well as on the
quantum level \cite{bohigas84}, it was natural to seek for a dynamical 
counterpart of 
Anderson localization in such a setting. Soon {\em dynamical
localization}, i.e., the quantum suppression of diffusive energy
growth in periodically driven, classically chaotic  
quantum systems was predicted \cite{fishman82,casati84,bluemel84a} for the
periodically kicked rotor and for microwave driven hydrogen atoms
initially prepared in highly excited Rydberg states. Whilst
experiments on cold atoms \cite{moore94} -- which closely mimic 
the kicked rotor in their center of mass motion --
have confirmed this prediction in much detail, the situation remained
controversial when it comes to the interpretation of experimental
results on Rydberg states of atomic hydrogen and of alkaline atoms
\cite{galvez88,bayfield89,breuer89c,fu90,arndt91,graham91,gallagher91,pmk91,fishman91,benson95}.  
There, the presence of
additional degrees of freedom, of the atomic continuum, and of a
multielectron atomic core prevent a direct mapping onto the Anderson picture, 
and considerably complicate the
unambiguous interpretation of experimental results. On the other hand,
such atomic systems obviously extend the potential realm 
of dynamical localization -- i.e. of Anderson localization on the
energy axis -- considerably, since they provide generic examples of energy
transport in light matter interaction.

In this Letter we present a unifying theoretical picture which, for the first
time,  
resolves all
apparent inconsistencies between experimental observations on
different atomic species under microwave driving
\cite{galvez88,bayfield89,fu90,arndt91,benson95,noel00}. On the basis of 
ample numerical data obtained from an accurate, approximation-free
treatment of the atomic excitation and ionization process, we identify a
scaling law under which the ionization thresholds of different atomic
species coalesce for sufficiently high driving
field frequencies, independently of their element specific,
unperturbed  spectral structure. This universality of the ionization
threshold provides strong support for the Anderson scenario as the
underlying transport mechnism, clearly beyond all evidence provided so
far. Furthermore, it allows a purely
spectral interpretation of the observed localization mechanism,
irrespective of the availability of a well defined classical
analogue.

Let us start with a condensed description of the physical system under study:
In a typical laboratory experiment \cite{galvez88,bayfield89,arndt91,noel00}, atomic one electron Rydberg states
prepared in an initial field free state $\rho_0 $ are exposed to a
classical monochromatic microwave field of frequency $\omega $ and field
strength $F$, for a duration $t$ of typically several hundreds to
several thousands of microwave periods $T=2\pi/\omega$. After the atom
field interaction, the percentage $P_{\rm surv}$ 
of atoms which did not ionize under
the external driving is registered as a function of $F$ and of the
initial principal quantum number $n_0$ which labels $\rho_0 $. At
fixed $\omega $ and $t$, the ionization threshold field $F(10\%)$
which induces $1-P_{\rm surv}=10\% $ is extracted from a series of
such measurements, over a large interval of $n_0$. For a microwave
frequency which is of the order of the local energy spacing $\Delta E$ 
in the unperturbed Rydberg series, approx. $N\simeq
1/2n_0^2\Delta E$ photons have to be absorbed by the atom to induce a
transition from the initial state $\rho_0 $ into the atomic continuum,
i.e. to produce a nonvanishing ionization signal. For typical values
of $n_0\simeq 25\ldots 100$, this implies multiphoton transitions of
the order $150\ldots 10$, which can mediate efficient ionization only
if they are composed of a sequence of near-resonant one photon
transitions between Rydberg states separated by $\omega\pm\delta $, 
with small detuning $\delta\ll \omega$. The increasing density of
states within the Rydberg progression towards the continuum threshold
guarantees the existence of such a sequence, provided $\omega $ is
large enough to depopulate $\rho_0 $ via a first near resonant one photon
transition. Under these premises, the intuitive analogy with the one
dimensional Anderson model is established by identifying the sequence
of near resonantly coupled Rydberg states with the neighbouring sites
of a tight binding potential, with transition matrix elements
essentially determined by the Rabi frequency which characterizes the
one photon coupling. The one photon coupling constants are effectively
randomized by the quasi random distribution of
detunings which is due to the nonlinearity of the Rydberg progression
to be set equal to $m\times\omega$, with integer $m$ \cite{brenner96}. On the
other 
hand, for $\Delta E\sim n_0^{-3}$ -- which defines the spacing of the
degenerate energy levels of atomic hydrogen -- the
quantum mechanical transport process along the energy axis can be
compared to classically chaotic transport in phase space, since
$\omega\sim  n_0^{-3}$ defines strong nonlinear coupling between the 
unperturbed classical Kepler motion of the Rydberg electron and the
external field, and hence leads to the efficient destruction of
invariant tori in phase space for sufficiently large values of $F$ \cite{casati84,bluemel84a}.
Given this classical picture of strongly driven Coulomb dynamics, the
results of typical experiments on atomic hydrogen are most often
represented in terms of scaled variables $\omega_0=\omega\times n_0^3$
and $F_0(10\%)=F(10\%)\times n_0^4$, what allows a direct identification of the
experimental (quantum) result with the associated classical phase
space structure which is completely determined by $\omega_0$ and
$F_0$. In such a plot, an increase of $F_0(10\%)$ is considered as a signature
of dynamical or Anderson localization, since classically diffusive/chaotic
ionization leads to a monotonous decrease of $F_0(10\%)$ with $\omega_0$, for
not too short atom-field interaction times \cite{benson95}.

Despite the fact that no uniquely defined classical one particle
dynamics is available for the excitation and ionization of
nonhydrogenic initial states of alkaline
Rydberg states, we compare in Fig.~1 the numerically generated 
scaled ionization thresholds of 
H, Li, Na, and Rb, as a function of $\omega_0$. 
\begin{figure*}[tbp]
\begin{center}
\centerline{\psfig{figure=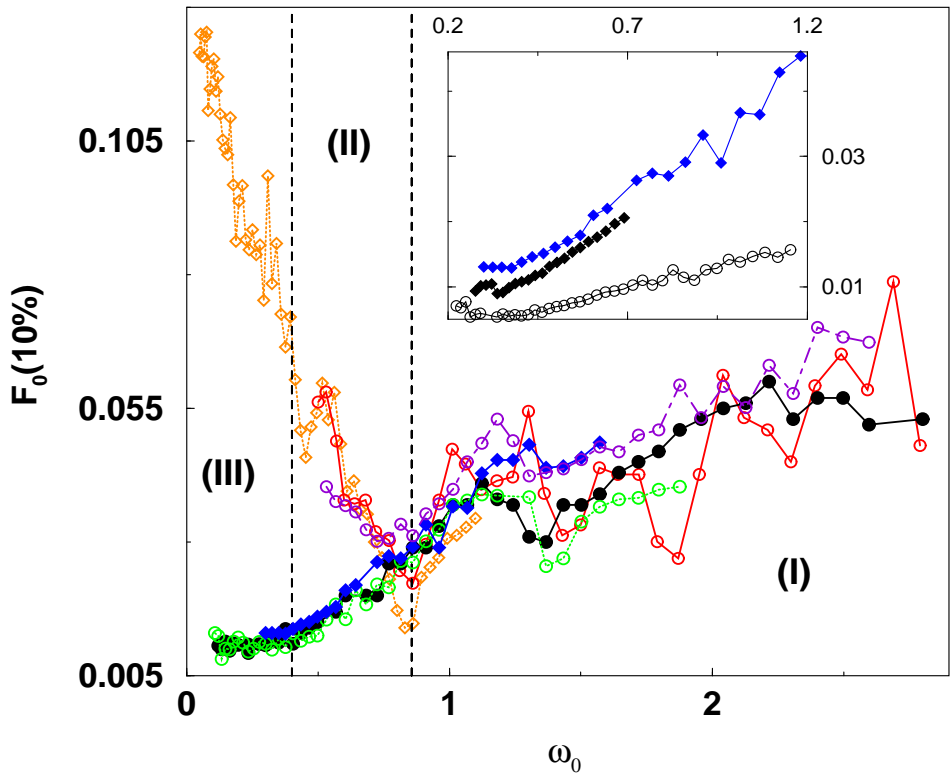,width=12.5cm,angle=-90}}
\end{center}
\caption{Ionization thresholds of H (open diamonds, dotted line, and open
  circles, full line: laboratory
  experiments \cite{galvez88,pmk95b}; open circles, dash-dotted line:
  numerical experiments), Li (filled circles, full line), Na (open circles,
  dotted line), and Rb (filled diamonds, full line), in scaled units
  $\omega_0$ and $F_0(10\%)$. All data were obtained for an interaction time
  $t=327\times T$, at driving frequency $\omega/2\pi =36\ \rm GHz$, with
  initial principal quantum numbers in the range $n_0=28\ldots 80$ except the
  ones represented by the open diamonds (here, $\omega/2\pi = 9.95\ \rm GHz$,
  and $n_0=32\ldots 90$ \protect\cite{pmk95b}). Hence, the comparison of
alkali and 
hydrogen 
  thresholds in scaled units does not imply any a priori assumption on the
  relvant atomic energy scales. Whilst all alkali data were obtained for {\em
  nonhydrogenic} initial states with angular momentum quantum numbers
  $\ell_0=0$ and $m_0=0$, we observe a species-independent, universal
  ionization threshold in the high frequency range (I) -- in strong support of
  the Anderson scenario as the dominant transport mechanism along the energy
axis. Inset: Comparison of scaled Rb ionization thresholds, for driving
  frequencies $\omega/2\pi=36\ \rm GHz$, $n_0=38\ldots 66$, $\ell_0=0$ (filled
  triangles) and
  $\omega/2\pi=8.867\ \rm GHz$, $n_0=59\ldots 80$, $\ell_0=1$ (filled
  pyramids), respectively, at the same interaction time $t=327\times
  T$. Clearly, the scaled thresholds essentially coincide, except for a
  systematic shift of the $8.867\ \rm GHz$ data to slightly smaller threshold
  values. This is a consequence of approaching the semiclassical limit
  $\hbar\sim n_0^{-1}\rightarrow 0$ whilst 
  keeping the scaled frequency fixed \protect\cite{pmk95b,abu95c}. Also shown
are 
  experimental data (open circles) obtained at $8.867\ \rm GHz$, though for
  much longer interaction times $t=44335\times T$ \cite{arndt91}. Whilst the
general 
  $\omega_0$ dependence is identical (notably the transition between frequency
  ranges (III) and (II)) for numerical and experimental data, the typically
  algebraic time dependence of the ionization threshold
  \protect\cite{arndt91,krugdiss} induces smaller thresholds for longer
  interaction times \protect\cite{abu95b,noel00}.}  
\label{Fig:tscan1}
\end{figure*}
The atoms are 
initially prepared in the low angular
momentum state $|n_0\, \ell_0=m_0=0\rangle$, with $\ell_0$ the
angular momentum and $m_0$ its projection on the polarization axis of
the linearly polarized driving field. These results were
obtained by diagonalization of the complex symmetric Floquet matrix
which represents the complex dilated Floquet Hamiltonian of the driven
atom, amended by the phase shift experienced by the electronic wave
function of the Rydberg electron upon scattering from the
multielectron core. Whilst details of our theoretical and numerical
treatment can be found elsewhere \cite{akab02,krugdiss}, let us just stress
here that the 
atomic object under study is treated without any approximations on its
dimensionality, that the atomic continuum is fully taken into account,
and that no adjustable parameters are available. The results for the
different species were obtained for fixed laboratory parameters 
$\omega/2\pi=36\ \rm GHz$ and $t=327\times T$, and over a broad range 
$28\leq n_0\leq 80$ of principal quantum numbers. Hence, representing
the data in scaled units does not imply any a priori assumption on the
relevant atomic energy scales, and the {\em only}
possible cause of different ionization thresholds of different species
is the element specific value of the nonvanishing quantum defects of
the low angular momentum states.  

Clearly, we can identify three regimes \cite{akab02} in Fig.~1, which are
defined by 
comparison of the hydrogen data with those for the nonhydrogenic
alkaline initial states. In regime (I), at large scaled frequencies
$\omega_0\geq 1$, all atomic species exhibit essentially {\em identical}
ionization thresholds, {\em irrespective} of the specific structure
of the unperturbed Rydberg spectra characterized by quite variable
quantum defects
ranging from $0$ to $3.6$, for $\ell\leq 3$. Only small relative
differences on top of the general trend of the ionization threshold in
this frequency regime reflect local spectral structures which differ
from species to species and which determine the weight of
individual atomic eigenstates in the field, in the decomposition of
the initial field free state
\cite{breuer89c,abu95b,krugdiss}. These differences are
particularly 
pronounced for the experimental hydrogen data \cite{galvez88,pmk95b} of the
Stony Brook group 
also 
shown in this plot, since in these experiments (which, in addition, 
start from a statistical mixture $\rho_0$ over the energy shell
labelled by $n_0$) the microwave is
switched on rather slowly, allowing for a complicated sequence of 
(a)diabatic transitions during the switching period of the microwave
field. Notwithstanding, the global agreement of numerical and
experimental data in regime (I) is rather impressive.

In regime (II), a strong discrepancy between the nonhydrogenic
alkaline thresholds and those of atomic hydrogen develops as we
proceed to smaller scaled frequencies, i.e. to smaller principal
quantum numbers at fixed laboratory frequency \cite{benson95,pmk95b}. 
This sharp transition
from (I) to (II) reflects the unavailability of a sequence of near
resonant one photon transitions out of the hydrogen initial state 
quite dramatically, since the hydrogen threshold can take values
larger by more than almost one order of magnitude as compared to the alkaline
threshold. Remarkably, the alkaline thresholds still coalesce in this
interval of principal quantum numbers, and the increase of
$F_0(10\%)$ with {\em increasing} $\omega_0$ shows that the
alkaline excitation and ionization process is still dominated by
Anderson localization in this interval of scaled frequencies. 

Only in regime (III) is the energy of the driving photon too small to
couple one photon transitions in the alkaline species, and the
threshold saturates accordingly (as a matter of fact, it slowly
increases with decreasing scaled frequency, what is however barely
realized on the scale of the figure -- see \cite{akab02} for an alternative
presentation of the lithium data).  

In the inset in Fig.~1, we compare the above ionization threshold 
of Rb at $\omega/2\pi
=36\ \rm GHz$ with the one at $\omega/2\pi= 8.867\ \rm GHz$, within the
same $\omega_0$ interval and accordingly chosen values of $n_0=59\ldots 80$, at
the same interaction time $t=327\times T$, and for a nonhydrogenic
initial state with $\ell_0=1$.
Clearly, 
both thresholds
conincide in scaled units, with a small systematic shift of the
$\omega/2\pi= 8.867\ \rm GHz$ results towards lower values, what is
simply due to the systematically larger values of $n_0$ imposed by the
frequency scaling for the smaller laboratory value of the driving
frequency \cite{abu95c}. Obviously, the classical scaling of frequency and
amplitude 
cannot account for the semiclassical limit -- which has to comply with
the correspondence principle and hence must reproduce the classical
ionization threshold which -- by definition -- is {\em smaller} than the 
quantum threshold in the Anderson/dynamically localized regime. 

The inset also contains experimental ionization thresholds of Rb
$\ell_0=1$ states under $\omega/2\pi= 8.867\ \rm GHz$ driving, though
for much longer interaction times $t=44335\times T$ \cite{arndt91}. 
Clearly, due to the
nonvanishing decay rate of the eigenstates of the atom in the field,
the experimentally observed ionization threshold is smaller than the
numerical one obtained for shorter interaction times, but the
transition between regimes (II) and (III) occurs in the same range at the same
of $\omega_0$ \cite{krugdiss}! Furthermore, we see that the time dependence of
the  
threshold is stronger for larger frequencies, thus slightly affecting
the slope of the threshold curve -- an observation also made in laboratory
experiments on Li \cite{noel00}. Indeed, a detailed study of the time
dependence of the ionization threshold reveals a generic, algebraic decay law
$F_0(10\%)\sim t^{-\gamma}$, where $\gamma$ tends to increase slowly with
$\omega_0$ 
\cite{krugdiss}. Whilst we could not reliably
access the experimental regime of interaction times, since this
requires extremely large a basis set such as to achieve numerical 
convergence of
the decay rates better than $10^{-13}\ \rm a.u.$, typical values of $\gamma$
are consistent with the gap opening up between the numerical and experimental 
$8.867\ \rm GHz$ data in Fig.~1 (note, however, that the precise value of
$\gamma$, at a given value of $n_0$ and $\omega$, will depend on the local
spectral structure and on the particular envelope of the microwave pulse
experienced by the atoms \cite{breuer89c,abu95b}, which our numerical approach
does  
not account for).  

Given our observations in Fig.~1, the following picture
emerges: Atomic one electron Rydberg states exhibit a universal
ionization threshold if the tight binding picture sketched initially
can be realized by a sequence of near resonant one photon transitions
connecting the initial field free state to the atomic continuum
\cite{brenner96}.  
Consequently, in regime (I), for $\omega_0\geq 1$, i.e. for photon
energies larger than the hydrogenic level spacing, 
alkaline atoms and atomic hydrogen exhibit
identical thresholds and ignore the additional nonhydrogenic level
structure provided by the nonvanishing quantum defects of the
different species. Also for alkaline atoms the scaling law
inherited from the classical Coulomb dynamics prevails, since the
energy spacing between the nonhydrogenic initial state
and the hydrogenic manifold exhibits the same dependence on the
principal quantum number as the hydrogenic energy
splitting \footnote{As a matter of fact, this scaling law also allows to
understand  
experimental data on Rb at $\omega/2\pi=12.6\ \rm GHz$ \cite{benson95}, where
a prominent enhancement of the ionization threshold was observed at $n_0=89$,
though could not be explained: In Coulomb scaling, this corresponds to a
scaled frequency $\omega_0\simeq 1.3$, and the local maximum of $F_0(10\%)$
thus is a remnant of the principal nonlinear resonance island in the classical
phase space of the driven two-body system \cite{pmk95b}.}. 
Nonhydrogenic states of alkali atoms exhibit
Anderson/dynamical localization in regime (II), in the absence of 
well defined classical one electron dynamics, simply due to the
availability of a sequence of near resonant one photon transitions to
the continuum. Note that in this regime this spectral substructure is
provided by the scattering of the electron from the atomic
multielectron core which, on the scale of a typical Rydberg orbit in
the energy range considered here, acts as a point scatterer
\cite{jonckheere98}. Given the 
tremendous enhancement of the alkaline ionization yield as compared to 
atomic hydrogen in this parameter range, this represents the arguably
most dramatic, directly observable signature of core scattering so far
observed in a chaotic quantum system: here, core scattering
restablishes Anderson localization! As compared to the core scattering
effect the nonvanishing time dependence of the ionization threshold
is but a correction: even increasing the interaction time by a factor 
$10\ldots 100$ does lower the threshold only weakly. All published experimental
data which were so far interpreted in terms of dynamical localization
were clearly obtained in regimes (II) and (III). The universal threshold
in regime (I) was a theoretical prediction \cite{krugdiss}, 
which, however, has been confirmed
in very recent laboratory experiments \cite{gallagher_bh03,maeda_prep04}
on one electron Rydberg states of Li and
Sr.    

Support as a Grand Challenge project at the Leibniz-Rechenzentrum of the
Bavarian Academy of Sciences is most gratefully acknowledged.

\bibliography{literatur04}
\end{document}